\documentclass[twoside]{dis09}
\usepackage[latin1]{inputenc}
\usepackage[dvips]{graphicx,epsfig,color}
\usepackage{wrapfig,rotating}
\usepackage{amssymb,amsmath,array}
\usepackage{hyperref}

\begin{document}
\title{Diffractive structure functions in nuclei}

\author{\underline{T.~Lappi}$^{1,2}$, H.~Kowalski$^{3}$, C.~Marquet$^{2,4}$ and R.~Venugopalan$^{5}$
%
%
\vspace{.3cm}\\
%
1- Department of Physics
P.O. Box 35, 40014 University of Jyv\"askyl\"a, Finland
\vspace{.1cm}\\
2- Institut de Physique Th\'eorique, CEA/DSM/Saclay,
91191 Gif-sur-Yvette, France
\vspace{.1cm}\\
3- Deutsches Elektronen-Synchrotron DESY,
22607 Hamburg, Germany
\vspace{.1cm}\\
4- Department of Physics, Columbia University,
New York, NY 10027, USA 
\vspace{.1cm}\\
5- Physics Department, Brookhaven National Laboratory,
Upton, NY 11973, USA
}

\maketitle

\begin{abstract}
We calculate proton and nuclear diffractive structure functions in the IPsat
(Kowalski-Teaney) dipole model.
This parametrization has previously been shown to provide good agreement
 with inclusive $F_2$ measurements and exclusive vector meson measurements at
 HERA.  We discuss how the impact parameter dependence crucially affects
 our analysis, in particular for small $\beta$.
\end{abstract}

\section{Introduction}
\nocite{dis09url}

The large fraction of diffractive events observed at HERA shows that 
modern colliders are approaching the nonlinear regime of QCD, where gluon 
saturation and unitarization effects become important. 
In Deep Inelastic Scattering (DIS) 
on nuclei the nonlinear effects are enhanced by the possibility of
interacting coherently with several
nucleons simultaneously~\cite{Kowalski:2007rw}. There are plans for several 
facilities capable of high energy nuclear DIS experiments, as the 
EIC~\cite{Deshpande:2005wd} and LHeC~\cite{Dainton:2006wd} colliders. 
Due to the difficulty in measuring an intact recoil nucleus deflected
by a small angle, diffractive eA collisions present an experimental 
challenge. But if they are successful, 
nuclear diffractive DIS (DDIS) would provide 
a good test of our understanding of high energy QCD.

In the high energy limit DIS is best understood in the dipole frame,
where the incoming virtual photon fluctuates into a quark-antiquark 
pair which then interacts with the target. The scattering
amplitude is related to the correlator of two Wilson lines in the
wavefunction of the nucleus. In contrast to the language
of collinearly factorized parton distribution functions,
in this formalism both inclusive
and diffractive observables can be calculated from the same
 universal dipole cross section. This enables one to naturally 
use the dipole cross sections fitted to one process to predict
observables in another one. In this talk we will review the
results of our recent work~\cite{Kowalski:2008sa} to apply
this ideology to computing nuclear diffractive
structure functions. Our emphasis is not on the 
most recent developments of high energy evolution equations,
but the effects of a more realistic and consistent impact parameter
dependence. This will lead
us to discuss, in addition to the consequences of 
nuclear geometry on diffractive observables, the importance
of the proton impact parameter profile used in the calculations.
In this paper we shall first describe the dipole cross sections 
and calculation methods used and then summarize our results
for nuclear DDIS.

\section{Method}
We decompose the diffractive structure function into different 
components in the standard way as
\begin{equation}
{x_\mathbb{P}} {F_2^{\textrm{D}}}({x_\mathbb{P}},\beta,Q^2)=
{x_\mathbb{P}} {{F_{T,q\bar{q}}^\textrm{D}}}({x_\mathbb{P}},\beta,Q^2) +
{x_\mathbb{P}} {{F_{L,q\bar{q}}^\textrm{D}}}({x_\mathbb{P}},\beta,Q^2)  +
{x_\mathbb{P}} {{F_{T,q\bar{q}g}^\textrm{D}}} ({x_\mathbb{P}},\beta,Q^2).
\end{equation}
For the lowest Fock state of the virtual photon wavefunction,
the $q\bar{q}$ dipole (${{F_{T,q\bar{q}}^\textrm{D}}}$ and 
${{F_{L,q\bar{q}}^\textrm{D}}}$ ) we follow the treatment 
of~\cite{Golec-Biernat:1999qd}. At small $\beta$ (large mass 
of the diffractive system) the dominant contribution comes from 
higher Fock states. In this work we are interested in the finite
experimentally relevant range of $\sqrt{s}$ and will only include
the leading (in ${\alpha_{\mathrm{s}}}$) $q\bar{q}g$-component of these. In different
works this component  has been evaluated in different limits,
we shall here use the approach of~\cite{Marquet:2007nf} and 
interpolate between the small $\beta$, large ${N_\mathrm{c}}$ formula
used in~\cite{Munier:2003zb} and the finite $\beta$, large $Q^2$
form used in~\cite{Golec-Biernat:1999qd}. We refer the reader to
the references above for the detailed formulae.

We use the ``IPsat'' dipole cross section parametrization introduced 
in Ref.~\cite{Kowalski:2003hm} and extensively studied 
in Ref.~\cite{Kowalski:2006hc}. This is an impact parameter dependent
dipole cross section that combines unitarization and 
the correct behavior of structure functions
 the logarithmic 
large $Q^2$ (i.e. small dipole size $r$) behavior of structure functions are achieved
This is achieved using an eikonalized DGLAP-evolved gluon distribution 
function~\cite{Bartels:2002cj}. The
dipole cross section is given by
\begin{equation}
{\frac{\, \mathrm{d} \sigma_\textrm{dip}}{\, \mathrm{d}^2 {\boldsymbol{b}_T}}}  
= 2\,\left[ 1 - \exp\left(- r^2  F(x,r) T({\boldsymbol{b}_T})\right) 
\right],
\label{eq:BEKW}
\end{equation}
where $F$ is proportional to the 
DGLAP evolved gluon distribution
\begin{equation}
F(x,r^2) = \frac{\pi^2}{ 2 {N_\mathrm{c}}} {\alpha_{\mathrm{s}}}  (\mu^2(r^2))
x g (x,\mu^2(r^2) ),
\end{equation}
with both the coupling and the gluon distribution $xg(x,Q^2)$
evaluated at the scale $\mu^2(r^2) = \mu_0^2 + C/r^2$.

\begin{wrapfigure}{r}{0.5\columnwidth}
\centerline{\includegraphics[width=0.45\columnwidth]{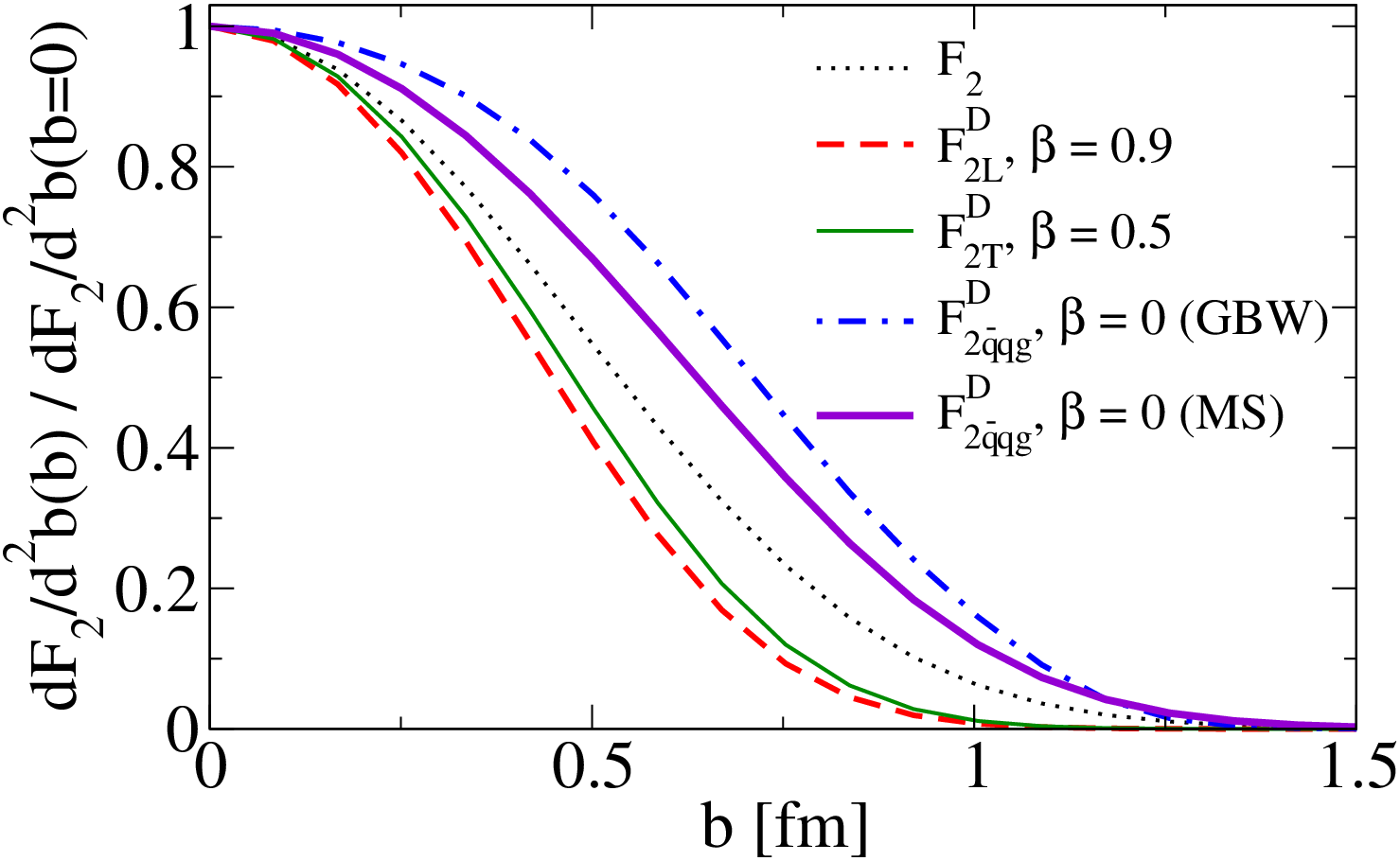}}
\caption{Contributions of different impact parameters to the inclusive and diffractive
structure functions in the proton in the IPsat model.}\label{fig:b}
\end{wrapfigure}
Several works on the subject (e.g.~\cite{Golec-Biernat:1998js,Golec-Biernat:1999qd,%
Iancu:2003ge,Kugeratski:2005ck})
assume, explicitly or implicitly, a factorizable ${\boldsymbol{b}_T}$ dependence
of the dipole cross section.
\begin{equation}\label{eq:factbt}
{\frac{\, \mathrm{d} \sigma_\textrm{dip}}{\, \mathrm{d}^2 {\boldsymbol{b}_T}}}
({\boldsymbol{b}_T},{\boldsymbol{r}_T},x) \sim e^{-\frac{{\boldsymbol{b}_T}^2}{2B}},
\end{equation}
which leads to an exactly exponential $t$-dependence of diffractive 
cross sections. The conceptual problem with the 
form Eq.~(\ref{eq:factbt}) is that it cannot be a solution 
of the BK equation  (unless the profile is a $\theta$ function, which would 
contradict the experimentally observed $t$-distribution).
A factorized Gaussian profile for the
proton dipole cross section, for example, \emph{does not approach the correct 
unitarity limit} for $b\neq0$. 
This is the main motivation for including the impact parameter
dependence in the saturation scale 
model~(\ref{eq:BEKW}), not as a factorizable prefactor of the
dipole cross section.
For a proton the impact parameter profile in Eq.~(\ref{eq:BEKW}) is taken as
$T({\boldsymbol{b}_T}) = T_p({\boldsymbol{b}_T})\sim\exp(-\frac{b^2}{2 B_{\rm G}})$ and for 
a nucleus $T({\boldsymbol{b}_T}) = \sum_{i=1}^A T_p({\boldsymbol{b}_T}-
{\boldsymbol{b}_T}_i)$, where 
the nucleon coordinates ${\boldsymbol{b}_T}_i$ are taken from a standard
Woods-Saxon distribution~\cite{DeJager:1987qc}.
The concrete consequence of this impact parameter dependence is that,
in contrast to the factorized ansatz~(\ref{eq:factbt}), the different 
components of the diffractive structure have different $b$-dependences
from each other and from the inclusive cross section. 
The $q\bar{q}$-component is enhanced at small $b$ (closer to the black disk limit),
whereas the $q\bar{q}g$-part is dominated by larger $b$ (because it
vanishes in the black disk limit). This structure is illustrated in 
Fig.~\ref{fig:b}.

\section{Results}

\subsection{HERA}

We compare our calculation to the HERA results on diffractive structure functions,
measured using both using the rapidity gap method (ZEUS FPC \cite{Chekanov:2005vv} and
H1 LRG \cite{Aktas:2006hy}) and by measuring the recoil proton 
(ZEUS LPS \cite{Chekanov:2004hy} and H1 FPS \cite{Aktas:2006hx}).
Because the FPC and LRG data include events in which the proton has broken up, 
the cross-sections measured for the process $ep\!\rightarrow\!eXY$ are larger 
than the one measured for the process $ep\!\rightarrow\!eXp.$
We scale down this data by a constant factor to correct for the
proton dissociation contribution;  the ZEUS FPC data by a factor
of 1.45 and the H1 LRG data by 1.23. These factors are different due to the
different cuts on $M_Y$, the mass of the proton dissociation system.
For the combined dataset from ZEUS and H1 data both with
 and without identified protons we get $\chi^2/N_\textrm{dof} = 1.3$, with
a coefficient ${\alpha_{\mathrm{s}}} = 0.14$ of the $q\bar{q}g$ term.
This is the values of ${\alpha_{\mathrm{s}}}$ that we shall use 
to evaluate nuclear diffractive structure  functions in the next section.
For the IPsat model
the largest contribution to the $\chi^2$ comes from the rapidity gap method data at 
large $\beta$. The fit to only the LPS 
($\chi^2 = 0.5$ IPsat) and FPS ($\chi^2 = 0.8$) is much better. 
Considering just the LPS also accommodates a larger value of ${\alpha_{\mathrm{s}}}=0.21$ with still 
$\chi^2<1$. 

\begin{figure}
\includegraphics[width=0.45\columnwidth]{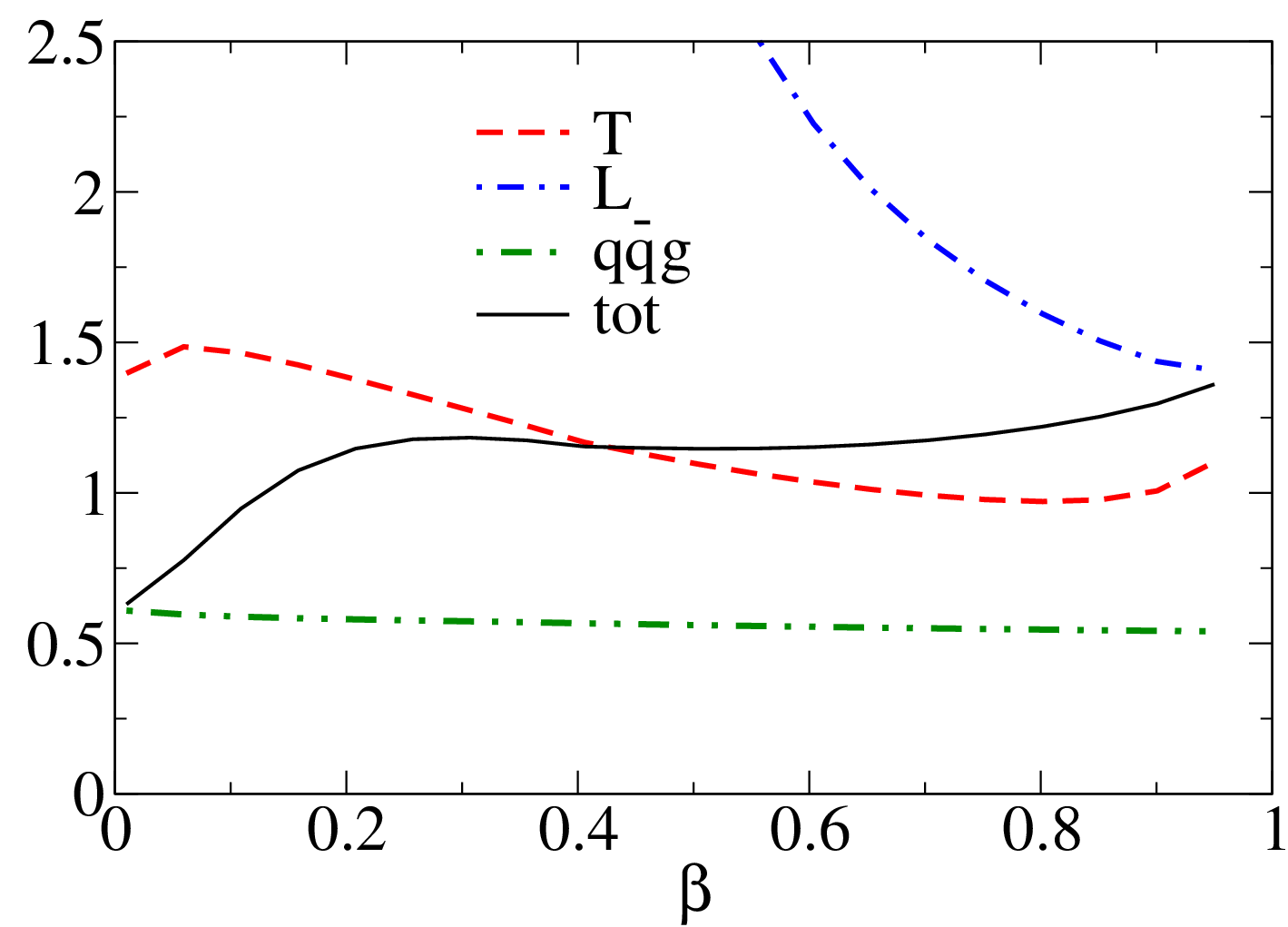}
\hfill
\includegraphics[width=0.45\columnwidth]{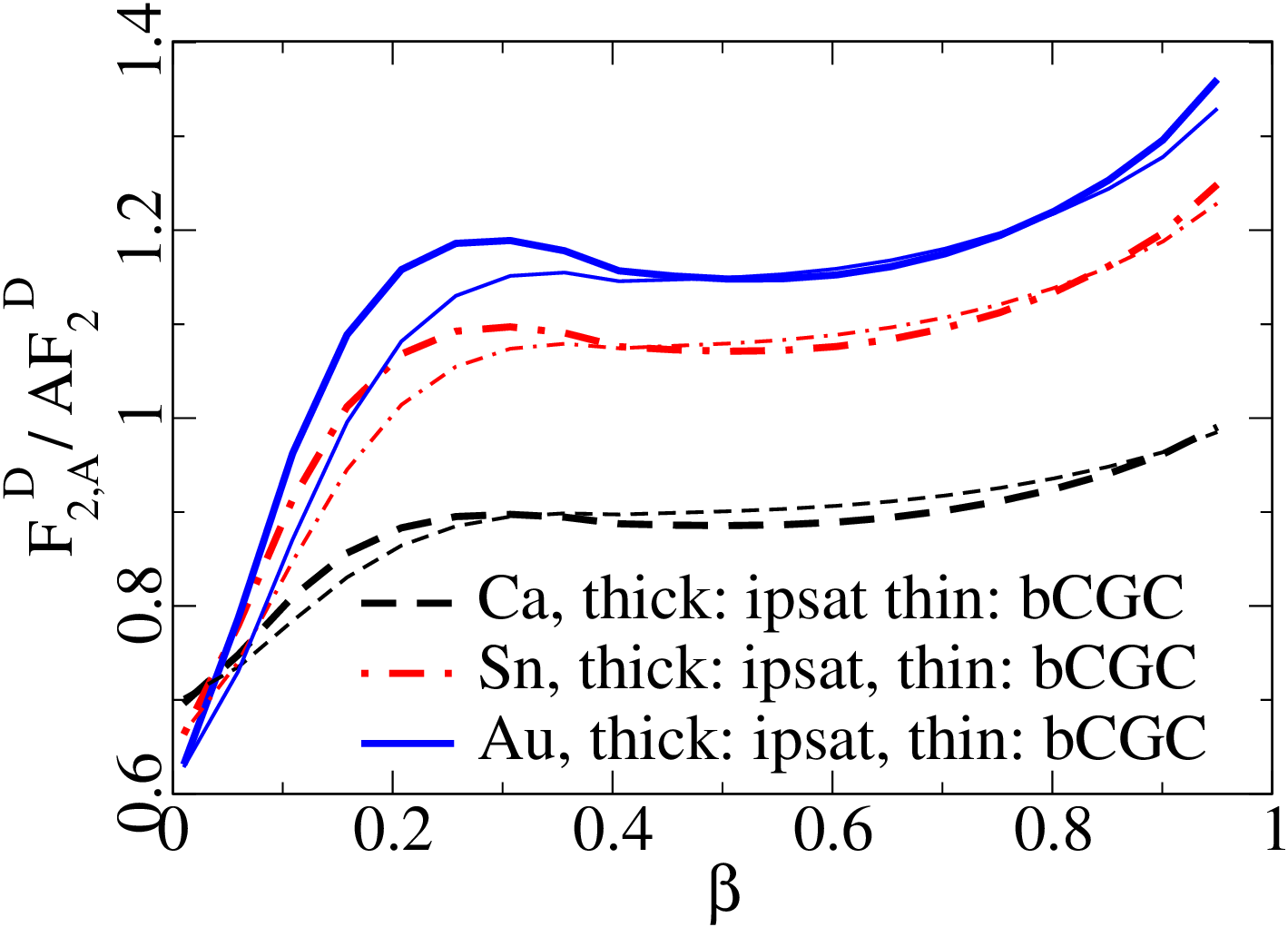}
\caption{
Left: dependence on  $\beta$ of nuclear effects on the individual components of
the diffractive structure function in a gold nucleus.
Right: total $\beta$-dependence of nuclear effects on the diffractive structure function.
In both plots $Q^2=5\textrm{ GeV}^2$ and ${x_\mathbb{P}}=10^{-3}$.
}\label{fig:betadep}\label{fig:totbetadep}
\end{figure}

The fit to HERA data is better with a smaller ${\alpha_{\mathrm{s}}}$ than in Ref.~\cite{Marquet:2007nf}.
Given the $b$-dependence described previously this is to be expected. The factorized 
$b$-dependence used in earlier calculations of the diffractive
structure function  such as Refs.~\cite{Golec-Biernat:1999qd,Marquet:2007nf}
forces the $q\bar{q}g$-component to have the same impact parameter dependence as
the $q\bar{q}$-component.
As discussed above, the $q\bar{q}g$ component
is sensitive to larger impact parameters and is thus larger; in order to fit the
same data this must be compensated by multiplying it with a smaller factor of ${\alpha_{\mathrm{s}}}$.

\subsection{Predictions for nuclei}

In Fig.~\ref{fig:betadep}  we show the ratios of different components of the gold 
diffractive structure function to the proton one as a function of $\beta$. 
The $q\bar{q}$-components of the ${{F_{2A}^\textrm{D}}}$ are enhanced compared to $A$
times the proton diffractive structure functions. This is to be expected, because of 
the fact that in a gold nucleus
the dipole cross section is, on average over the transverse area, closer to the 
unitarity limit than the proton - it is ``blacker''. The elastic 
scattering probability of a
$q\bar{q}$ dipole is maximal in the ``black disk'' limit and the approach 
to it is quicker in a large nucleus. 
The $q\bar{q}g$ component, on the other hand, is suppressed for nuclei compared to the proton. 
This is due to the fact that in a nucleus the scattering amplitude is closer to the unitarity
limit, when the $q\bar{q}g$ component vanishes.
 This leads to a nuclear suppression of the diffractive structure function in the 
small $\beta$ region,  where the $q\bar{q}g$ component dominates.
The net result of the different contributions is that ${{F_{2A}^\textrm{D}}}$, for a large range in $\beta$,
is close to $A {{F_{2p}^\textrm{D}}}$.
In Fig.~\ref{fig:totbetadep}, we plot the total ratio as a function of $\beta$ for 
different nuclei  in the ``non breakup'' case. As expected from our prior
 discussion, one sees a strong enhancement with $A$ for larger $\beta$ and likewise, a 
stronger suppression with $A$ at very small values of $\beta$. 

\begin{figure}
\includegraphics[width=0.45\textwidth]{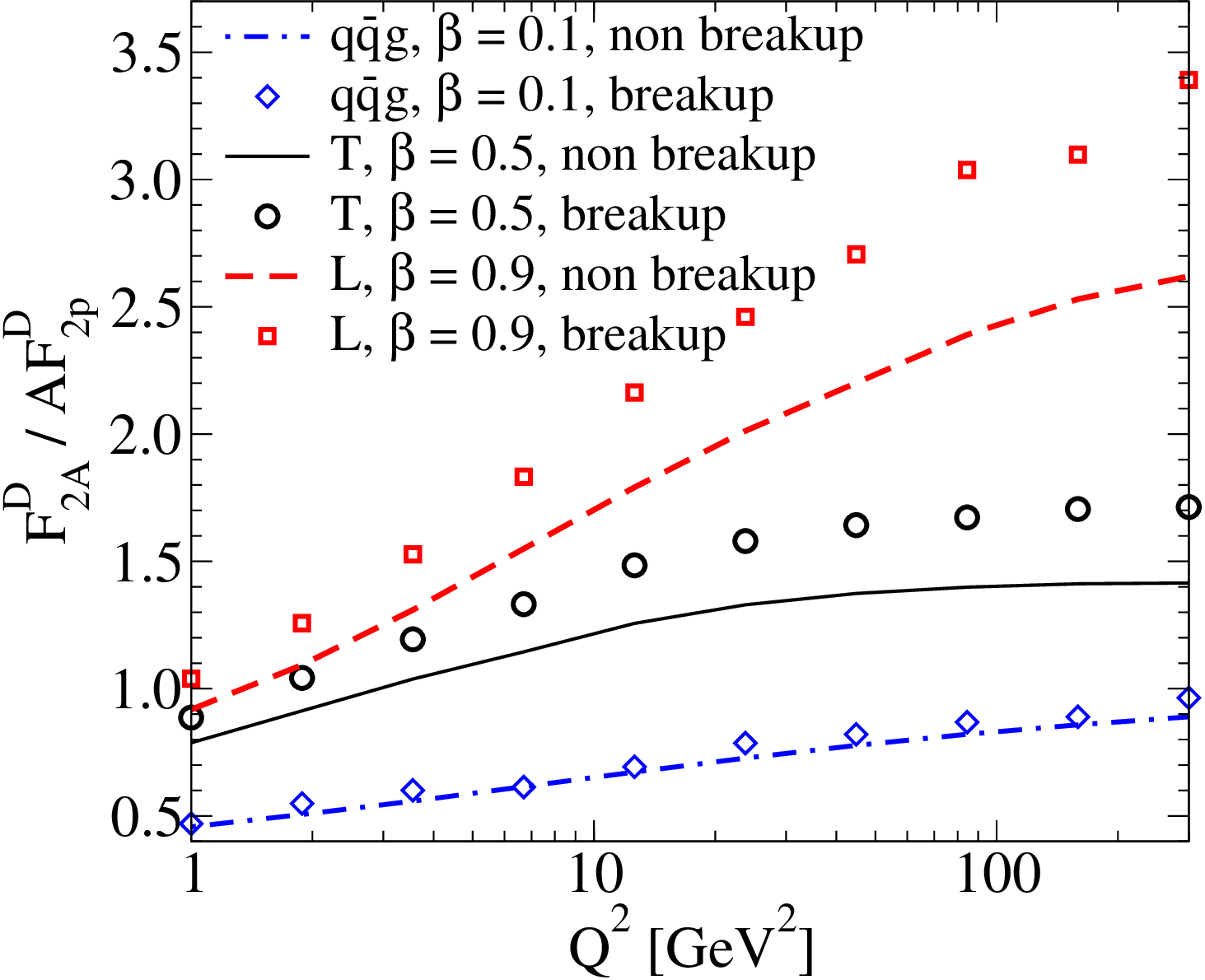}
\hfill
\includegraphics[width=0.48144\textwidth]{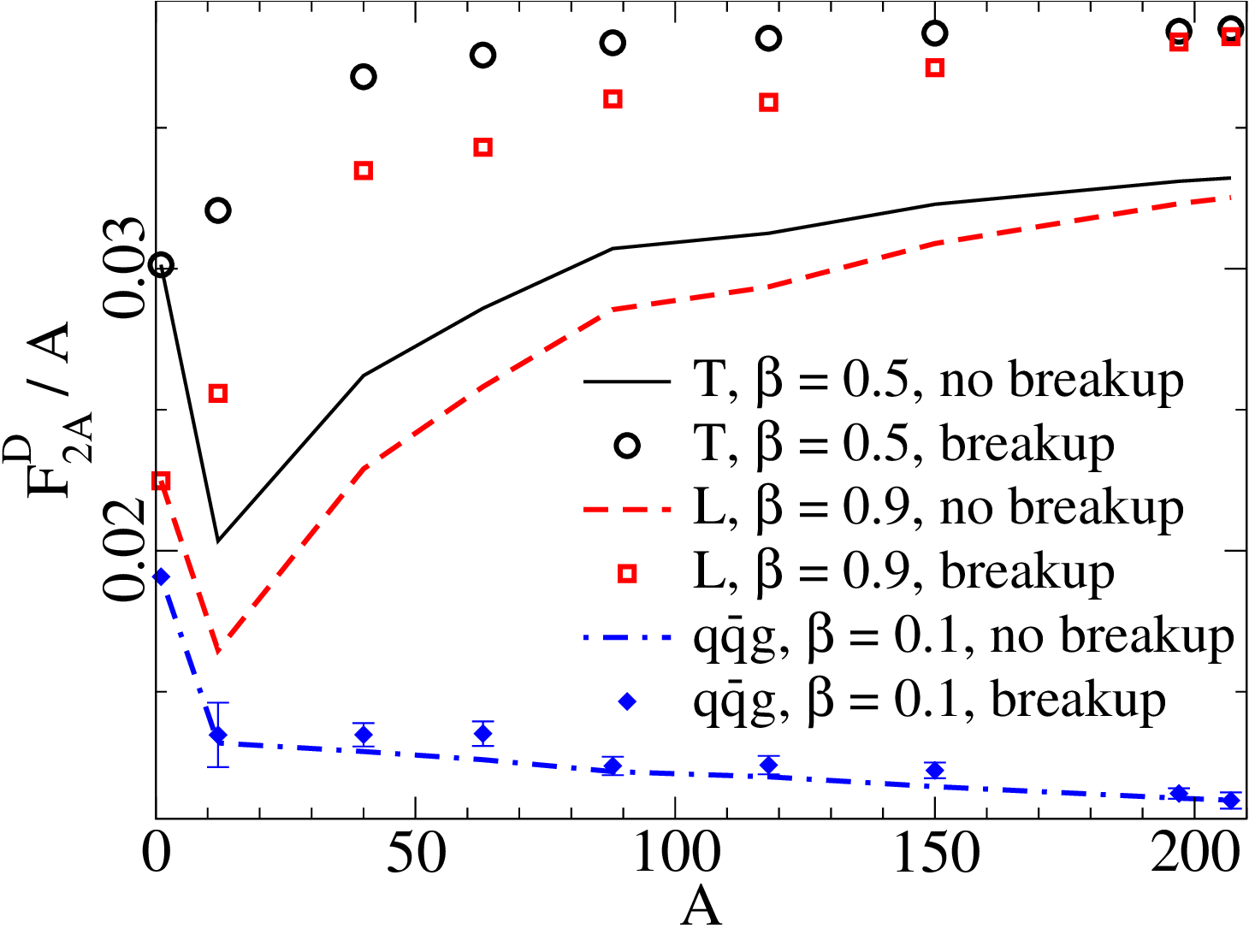}
\caption{
Left: dependence of the gold diffractive structure function on $Q^2$. 
Right: dependence of the nuclear diffractive structure function on the mass number $A$
for $Q^2=5\textrm{ GeV}^2$.
In both plots ${x_\mathbb{P}}=10^{-3}$.
}\label{fig:Qsqrdep}\label{fig:Adep}
\end{figure}

In our formalism, if one requires that the nucleus stays completely intact, the 
average over the nucleon positions ${\boldsymbol{b}_T}_i$ must be performed at the 
amplitude 
level. This is the case of \emph{coherent} diffraction. If the
nucleus is allowed to break up into color neutral constituents
(referred to as \emph{incoherent} diffraction), the average is performed at the level
of the cross section.  Measuring the 
intact recoil nucleus at such a small $t$ experimentally 
at a future electron ion collider
is challenging, so it is useful to consider both cases.
A comparison of the ``breakup'' 
versus ``non breakup'' cross-sections can be seen in the left panel of
Fig.~\ref{fig:Qsqrdep} for the  ratio of 
diffractive cross-sections as a function of $Q^2$.
The results in Fig.~\ref{fig:Qsqrdep} for the ratio of diffractive structure 
functions indicate that the diffractive cross-section in nuclei 
decrease more slowly for large $Q^2$ than in the proton. This can be understood
as a consequence of $Q_\textrm{s}$ being larger for nuclei and diffraction
being much more sensitive to $Q^2/Q_\textrm{s}^2$ than inclusive DIS.
In the right panel of Fig.~\ref{fig:Adep}, the nuclear size $A$ 
dependence of the longitudinal 
and transverse components of the diffractive structure function is 
shown for the ``breakup'' and ``non breakup'' cases. In the ``breakup'' case, one 
sees a very weak $A$ dependence.
In the coherent ``non breakup'' case, one first notes that the 
diffractive structure function first decreases up to atomic numbers $A\sim 10$, 
before beginning to rise.  As noted in 
Ref.~\cite{Kowalski:2007rw}, this is due to the typical scattering amplitude
for small nuclei actually being smaller than for a proton because of the diluteness 
of the nucleus. This leads to a suppression of coherent diffraction. 
The ``breakup'' case, on the other hand, can only be enhanced in nuclei.
For gold nuclei, the cross sections in the 
``non breakup'' case are about $15$\% lower than in the ``breakup'' case.

Because of the different nuclear modifications in inclusive and diffractive scattering,
the fraction of diffractive events in an experiment
depends on the detailed kinematics and experimental coverage.
For moderate values of
$Q^2$ and large nuclei we expect a nuclear shadowing of the inclusive structure
function by a factor $\sim 0.8$~\cite{Kowalski:2007rw}. A typical nuclear 
enhancement of diffraction (at moderate values of $\beta \gtrsim 0.2$)
is a factor of $\sim 1.2$. Combining
these we expect $\sigma_\textrm{D}/\sigma_\textrm{tot}$
 to be increased by a factor of $1.2/0.8 = 1.5$ compared to the 
proton. Thus from a typical ep fraction of 15\% we expect 
$\sigma_\textrm{D}/\sigma_\textrm{tot}$ to go up 
to 20\% -- 25\% at an eA collider.

\section*{Acknowledgments}
T. Lappi is supported by the Academy of Finland, project 126604.
RV's research is supported by DOE Contract No. DE-AC02-98CH10886.
CM's research is supported by the European Commission under the FP6
   program, contract No. MOIF-CT-2006-039860.


\begin{footnotesize}

\bibliographystyle{h-physrev4mod2}
\bibliography{spires}

\end{footnotesize}


\end{document}